\documentclass[conference]{IEEEtran}
\IEEEoverridecommandlockouts

\usepackage{cite}
\usepackage{amsmath,amssymb,amsfonts}
\usepackage{algorithmic}
\usepackage{graphicx}
\usepackage{textcomp}
\usepackage{xcolor}
\usepackage{comment}
\usepackage{url}
\usepackage{booktabs}
\usepackage{algorithmic}
\usepackage{algorithm}
    
\usepackage[pscoord]{eso-pic} 

\begin{document}

\title{Cross-Cluster Networking to Support Extended Reality Services}

\author{
\IEEEauthorblockN{Theodoros Theodoropoulos\IEEEauthorrefmark{1}\IEEEauthorrefmark{2},~Luis Rosa\IEEEauthorrefmark{3},~Abderrahmane Boudi\IEEEauthorrefmark{4},~Tarik Zakaria Benmerar\IEEEauthorrefmark{4},~Antonios Makris\IEEEauthorrefmark{1}\IEEEauthorrefmark{2}\\,~Tarik Taleb\IEEEauthorrefmark{5}\IEEEauthorrefmark{6},~Luis Cordeiro\IEEEauthorrefmark{2},~Konstantinos Tserpes\IEEEauthorrefmark{1}\IEEEauthorrefmark{2},and~JaeSeung Song\IEEEauthorrefmark{6}}

\IEEEauthorblockA{\IEEEauthorrefmark{1}\textit{Harokopio University of Athens}
}
\IEEEauthorblockA{\IEEEauthorrefmark{2}\textit{National Technical University of Athens}
}
\IEEEauthorblockA{\IEEEauthorrefmark{3}\textit{OneSource}
}

\IEEEauthorblockA{\IEEEauthorrefmark{4}\textit{ICTFICIAL OY}
}

\IEEEauthorblockA{\IEEEauthorrefmark{5}\textit{Oulu University}
}

\IEEEauthorblockA{\IEEEauthorrefmark{6}\textit{Sejong University}
}

}


\maketitle

\begin{abstract}
Extented Reality (XR) refers to a class of contemporary services that are intertwined with a plethora of rather demanding Quality of Service (QoS) and functional requirements. Despite Kubernetes being the de-facto standard in terms of deploying and managing contemporary containerized microservices, it lacks adequate support for cross-cluster networking, hindering service-to-service communication across diverse cloud domains. Although there are tools that may be leveraged alongside Kubernetes in order to establish multi-cluster deployments, each one of them comes with its drawbacks and limitations. The purpose of this article is to explore the various potential technologies that may facilitate multi-cluster deployments and to propose how they may be leveraged to provide a cross-cluster connectivity solution that caters to the intricacies of XR services. The proposed solution is based on the use of two open source frameworks, namely Cluster API for multi-cluster management, and Liqo for multi-cluster interconnectivity. The efficiency of this approach is evaluated in the context of two experiments. This work is the first attempt at proposing a solution for supporting multi-cluster deployments in a manner that is aligned with the requirements of XR services. 

\end{abstract}

\begin{IEEEkeywords}
 Kubernetes, Cloud, Edge, Continuum, Cluster API, Liquid Computing, Liqo, Network, XR, Immersive Services, 5G, and 6G.
\end{IEEEkeywords}

\section{Introduction}

Contemporary applications are deployed in the form of containerized microservices. To facilitate the underlying orchestration complexity of containerized microservice deployments, the notion of container orchestration frameworks was introduced. Kubernetes \cite{burns2022kubernetes} is an extensible open-source orchestration platform for automating software deployment, scaling, and management of containerised workloads and services and is considered to be the standardized way of orchestrating containers and deploying distributed applications. While Kubernetes is extremely popular in cloud computing environments, lightweight versions, such as K3S, are often deployed in Edge computing environments.

Furthermore, contemporary applications, such as eXtended Reality (XR), are often intertwined with a plethora of demanding Quality of Service (QoS) and functional requirements. The backbone of XR applications relies on providing an immersive experience for the various end-users. Providing acceptable levels of immersion requires extremely low latencies and high bandwidths. The scientific literature has showcased that for an end-user experience to be considered satisfactory, the end-to-end latency shall not be greater than 15ms, and the available bandwidth should be scalable up to 30 Gbps\cite{10.1145/2938559.2938583}. Furthermore, XR applications are extremely demanding in terms of computational resources since they incorporate functionalities such as the rendering of complex 3D models and the use of highly-defined graphics.

Another important requirement of XR services is the need for end-user equipment to be as light-weight and inexpensive as possible. While cloud computing can shift the computational adequacy burden to various remote resources, thus allowing end-user devices to be mobile and cost-effective, it cannot fully support immersive applications that require low latency and high bandwidth, since the end-user devices are usually far from the cloud servers. This fact leads to processing and network overheads, thus resulting in low performance and high latency. Edge computing aims at reducing the amount of data that needs to be transmitted to remote clouds and allows for data processing near the data sources. Thus, edge computing can provide faster response times, higher transfer rates, and better scalability and availability. Consequently, running XR services in a distributed fashion across the cloud-edge fabric would benefit application developers and help keep up with the aforementioned QoS requirements \cite{taleb2022towards}.

\begin{figure*}
    \centering
    \includegraphics[width=\linewidth]{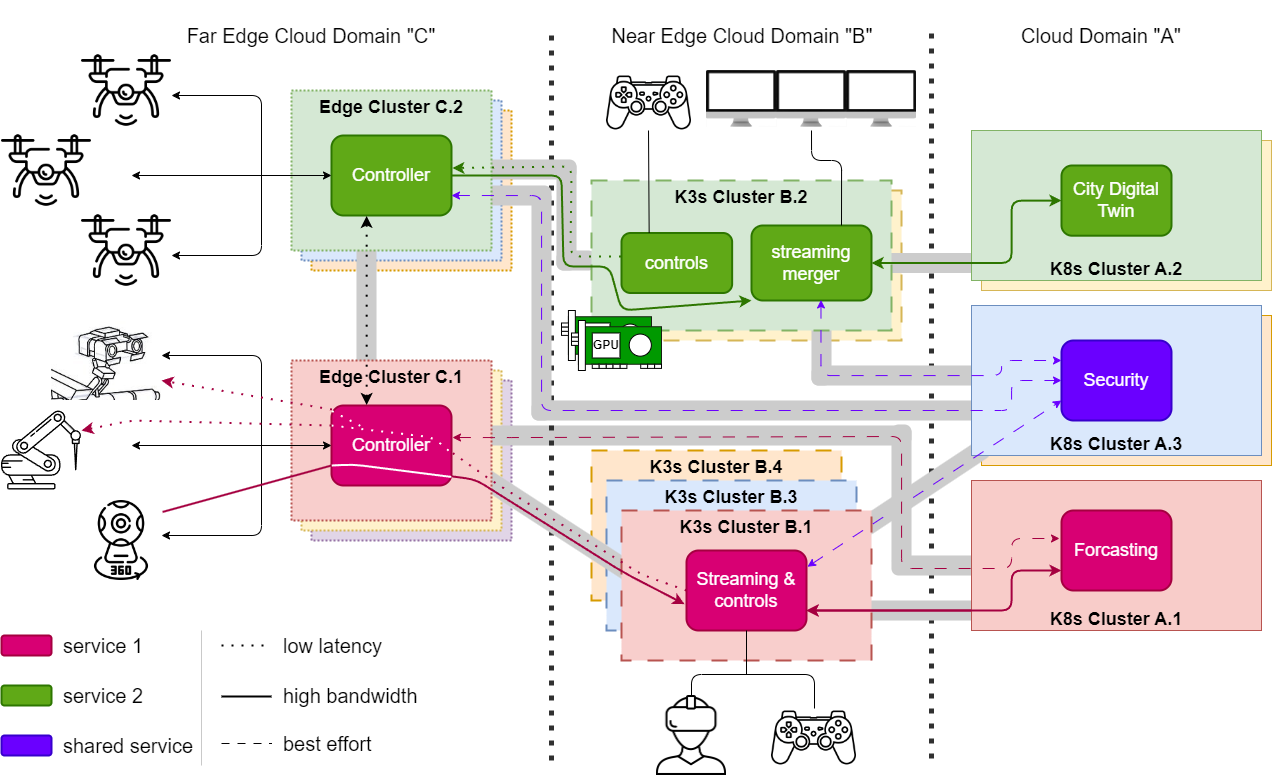}
    \caption{Multi-Cluster Example Use Case.}
    \label{fig:usecase}
\end{figure*}

Such a XR service deployment scenario that spans across the cloud-edge fabric is depicted in Fig.~\ref{fig:usecase}, whereby some distant users must collaborate in a virtual reality environment and would require many services to run concurrently over a wide area. As stated before, XR services require extremely low latency, which means that part of the service must be deployed close to stakeholders (i.e., domains B and C in Fig.~\ref{fig:usecase}). In addition, by having many users collaborating closely, communications paths need to be spun between the deployed services in a peer-to-peer fashion; whilst also communicating with some carefully placed shared services, e.g., used for security or synchronization purposes. Other resource-intensive services can also be placed in the cloud. It is clear that for this relatively simple, yet realistic, XR service, and with the many shifting requirements of the sub-services and their placement, leveraging cloud-edge deployments would greatly support the provisioning of such services.

Although it is technically feasible for a single Kubernetes cluster to span across various edge \& cloud sites that may be at vastly different locations, multi-cluster deployments offer numerous advantages. These include the ability to shift workloads between clusters to prevent congestion or failure, minimizing downtimes by seamlessly transferring workloads to alternative clusters, overcoming size constraints of single-cluster deployments, ensuring compliance with country/region-specific regulations for data storage in cloud-based workloads, allowing service providers flexibility in choosing algorithms and vendors, reducing costs, and enabling services to be distributed across cloud, edge, and proprietary infrastructure to avoid vendor lock-in. This approach facilitates continuous service optimization based on user satisfaction, running costs, and energy consumption. To ensure seamless cross-cluster communications in multi-cluster deployments, attention is needed for both cluster management and connectivity between clusters.

Unfortunately, contemporary versions of Kubernetes are unable to facilitate multi-cluster deployments. This limitation extends to two fronts, the first one of which is the orchestration and management of the multiple clusters that facilitate the various XR services. The second one is the communication among these XR services, across cluster boundaries. This endeavour becomes quite challenging when considering that XR services, on top of the aforementioned QoS requirements, are characterized by functional requirements, such as the need for User Datagram Protocol (UDP) support that serves as the cornerstone for numerous streaming use-cases, which are of paramount importance in the frame of XR services \cite{chung2022xr}. Furthermore, edge clusters are usually limited in terms of computational resources, and thus any attempt at implementing cross-cluster networking in a manner that spans across the cloud-edge fabric should take resource consumption into consideration (especially in the case of resource-intensive services, such as XR). Finally, to fully harvest the benefits of multi-cluster deployments, an ideal solution should facilitate dynamic, multi-ownership deployment scenarios and have a singular cross-cluster control plane to optimally schedule the various workloads. 

Towards achieving this goal, this paper is dedicated to examining numerous state-of-the-art multi-cluster connectivity \& management frameworks, and to proposing a novel solution that caters to the intricacies of XR services. More specifically, the proposed solution consists of two contemporary frameworks whose modus operandi is based on the support of multi-cluster deployments. These frameworks are \textit{Cluster API} for deploying and managing multiple clusters, and \textit{Liqo} for cross-cluster internetworking.

\section{Multi-Cluster Management}
\label{dep}
Multi-cluster management is a critical aspect of modern computing infrastructure, enabling organizations to efficiently oversee and coordinate multiple clusters of resources. 

\textbf{KubeFed} \cite{fed} is a multi-cluster management framework that enables each cluster to leverage its own local master. Unfortunately, this approach comes at the cost of additional software resources to support localised autonomy and synchronization across the various clusters. Furthermore, KubeFed supports multi-vendor environments, but the level of support may vary depending on the specific vendor and their level of compatibility with the Kubernetes API. 

\textbf{Karmada} \cite{kar}, on the other hand, is based on a different approach that utilizes a custom API Server. This server operates as a centralized control point and resembles the standard Kubernetes API, while high-level resources like Deployments are handled by custom controllers instead of following the standard workflow. This is determined by policy constraints set through Custom Resource Definitions. However, this approach does not fully adhere to Kubernetes, so administrators cannot manage lower-level objects such as pods in the context of effective monitoring and debugging.

\textbf{Terraform} \cite{terra}, a cloud-agnostic infrastructure provisioning tool, allows the creation of resources from various cloud services using a unified infrastructure-as-code approach. Despite its declarative approach for defining the desired end-state of infrastructure, the use of a single end-state file can lead to performance issues, particularly in multi-cluster deployments outside the same network. Additionally, unlike other explored multi-cluster management tools, Terraform is not free, which could be a deterrent for smaller application developer groups.

\textbf{Cluster API} \cite{cluster} is a free and open-source framework that brings Kubernetes-style APIs and support for the lifecycle management of (workload) Kubernetes clusters. Cluster API does not restrict to a specific infrastructure vendor, but lies on the concept of accommodating different cloud providers. This enables consistency, portability, automation and repeatability to cluster deployments, ultimately widening the possibility of orchestrating heterogeneous and multi-cloud domains in a unified and vendor-neutral fashion.

\section{Multi-Cluster Interconnectivity}
\label{con}

\begin{figure}
    \centering
    \includegraphics[width=\linewidth]{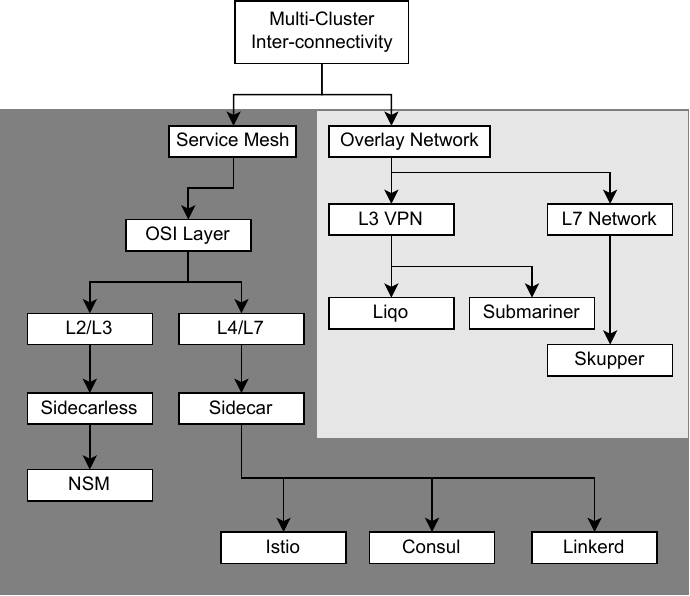}
    \caption{Multi-Cluster Interconnectivity Taxonomy.}
    \label{fig:taxonomy1}
\end{figure}

While the aforementioned frameworks are capable of establishing multi-cluster management, they do not provide any form of multi-cluster interconnectivity functionalities. Thankfully, various cloud-native solutions have emerged to provide such functionalities. Fig.~\ref{fig:taxonomy1} depicts an exhaustive taxonomy of tools enabling interconnectivity across \textit{Kubernetes} clusters. This taxonomy mainly focuses on two categories: \textit{Service Mesh} based approaches with multi-cluster support and \textit{Overlay Networks}. The range of enabling tools is selected based on a combination of their perceived maturity and documentation quality, empirical experience conducted by the authors of this article and open-source version. All of the following frameworks are capable of achieving multi-cluster connectivity. However, the ideal solution should be also capable of fulfilling the following requirements that are associated with XR services:
\begin{itemize}
    \item to not significantly increase resource consumption,
    \item to not contribute towards significantly increasing end-to-end latency,
    \item to support UDP,
    \item to have a singular cross-cluster control plane for optimal workload scheduling, and 
    \item to facilitate dynamic, multi-ownership deployment scenarios.
\end{itemize}

\subsection{Service Mesh Solutions}
    
A service mesh is a dedicated network infrastructure layer that manages communication between services in an application. It manages service requests, providing features like service discovery, load balancing, encryption, and failover. A service mesh typically relies on proxies (so-called sidecars) to form a mesh at the component level or, for instance, the host level.
    
\textbf{Istio} \cite{istio}, \textbf{Linkerd} \cite{link} and \textbf{Consul} \cite{con} are service mesh platforms designed for microservices integration, traffic management, policy enforcement, and telemetry data aggregation. They employ Sidecar proxies to handle traffic between services within a cluster, forming a microservice mesh offering service discovery, Layer~7 routing, circuit breakers, policy enforcement, and telemetry recording. However, a notable limitation is that, operating at Layer~7, they lack support for UDP traffic, a critical aspect for XR services that heavily rely on UDP for operations like video streaming. While a potential solution, such as CONNECT UDP, has been proposed, there is currently no practical option for supporting UDP communication in Layer~7 service meshes. Additionally, service mesh solutions with sidecar proxies introduce high application and latency overhead, which can be prohibitive for latency-sensitive applications.

The aforementioned service mesh solutions are capable of tackling challenges related to Layer~7 networking but are insufficient to facilitate use cases requiring Layer~2/3 networking. On the contrary, \textbf{Network Service Mesh} (NSM) \cite{nsm} offers a cloud-native network solution for microservices-based applications, emphasizing programmable and intelligent networking for Layer~2/3 connectivity. However, similar to other service mesh solutions, NSM lacks a cross-cluster control plane, preventing dynamic workload scheduling across diverse locations for optimal performance.

\subsection{Overlay Network Solutions}
An overlay network is a virtual (or logical) network that is established on top of an existing physical network. All nodes in an overlay network are connected with one another using virtual (or logical) links. Each link corresponds to a specified path in the underlying network topology. Overlay Network approaches use peering strategies (e.g. using VPN tunnels) to automatically interconnect clusters (and services). The code idea behind overlay networks is the facilitation of communications among micro-services regardless of their locations in a multi-cloud environment through the overlay network. Whether for federation or edge-cloud scenarios, these dynamic network topologies can also be helpful as a building block that does not enforce a specific orchestration platform but allows seamless communication across multiple sites.

\textbf{Submariner} \cite{sub} is a multi-cluster networking solution designed for cloud-native applications, offering seamless connectivity between Kubernetes clusters across various locations. It provides key functionalities such as Cross-cluster Layer~3 connectivity through encrypted VPN tunnels, service discovery, and network policy enforcement, enabling communication between services in different clusters, even across different cloud providers or data centers. While Submariner excels in achieving cross-cluster connectivity, it does not address workload orchestration and observability, leaving these aspects to static approaches or external tools.

\textbf{Skupper} \cite{skup} utilizes Virtual Application Networks (VANs) to address multi-cluster communication challenges, creating virtual networks connecting applications and services in a hybrid cloud at Layer~7. In Kubernetes, Skupper forms a network with each namespace having a Skupper instance, constantly sharing information about exposed services to create awareness across instances. Through annotation, Skupper exposes services, creating proxy endpoints available in all network namespaces, and, like many service mesh solutions, operates on Layer 7 networking, lacking support for UDP.

\textbf{Liqo} \cite{9984946} is a sophisticated open-source framework designed to enable seamless connectivity among clusters distributed across various geographical locations, encompassing on-premises environments, edge devices, and cloud infrastructures. Operating on a peer-to-peer model, Liqo establishes secure, encrypted connections between clusters to verify their identities. The architecture leverages a virtualization approach where remote clusters are abstracted as virtual nodes within the local cluster. This abstraction allows for transparent communication between interconnected clusters, regardless of the underlying Container Network Interface (CNI) plugin. In the context of bidirectional peerings, Liqo creates virtual nodes in each cluster, serving as representations of the resources provided by the remote cluster. 

Furthermore, Liqo introduces the concept of offloading, enabling the reflection and execution of workloads such as namespaces, services, and pods on these virtual nodes. This capability facilitates the exposure of services and execution of workloads in remote clusters. For instance, when a namespace is offloaded, Liqo dynamically creates a twin namespace in the remote cluster. This twin namespace allows pods and services to run seamlessly in a shared, cross-cluster environment. In the pod offloading scenario, the actual execution of pods and associated services is moved to a peered cluster, demonstrating the flexibility to optimize resource usage across clusters. This is particularly useful for demanding computing tasks like video processing, enabling efficient workload distribution based on the capabilities of different clusters. On the other hand, service offloading involves exposing only Kubernetes services on a remote cluster while retaining pod execution in the original cluster. This strategy allows for more selective offloading, giving flexibility in optimizing specific components of the application. 

Out of all the explored multi-cluster connectivity solutions, only Liqo manages to satisfy all of the functional requirements that are associated with XR services in terms of providing support for UDP, establishing a singular cross-cluster control plane for optimal workload scheduling, and facilitating dynamic, multi-ownership deployment scenarios. In section 

\section{Proposed Solution}
\label{proposed}
The proposed solution for establishing cross-cluster networking to support XR services consists of two contemporary frameworks that are capable of supporting multi-cluster deployments. These frameworks are Cluster API for deploying and managing multiple clusters, and Liqo for cross-cluster internetworking. To the best of our knowledge, this combination is novel and unexplored from both the design and implementation viewpoints. 

\begin{figure}
  \centering
  \begin{tabular}{ c @{\hspace{30pt}} c }
    \includegraphics[width=\columnwidth]{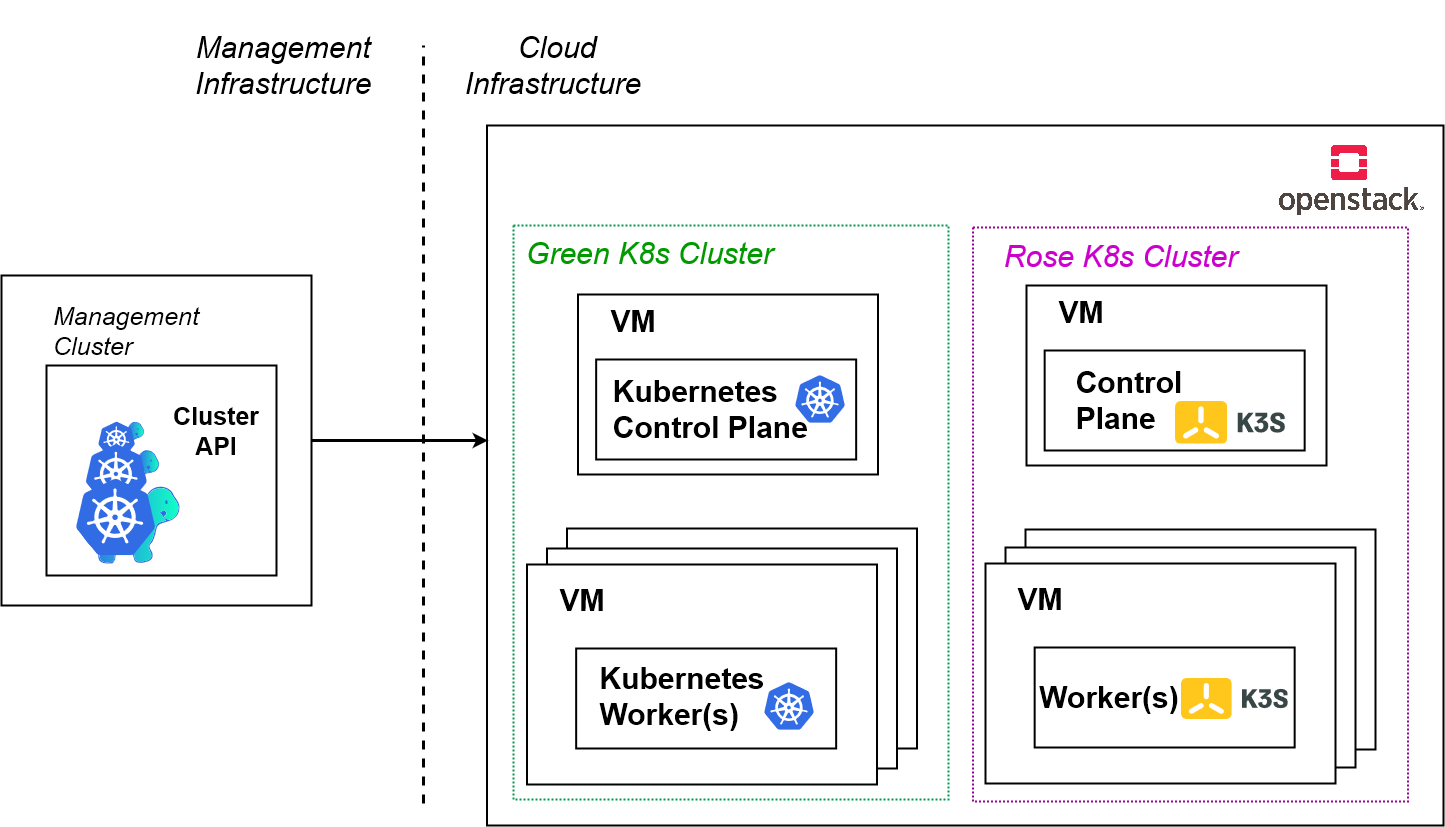} \\
    \small (a) Multi-Cluster Provisioning Scenario with Cluster API. \\ \\
    \includegraphics[width=\columnwidth]{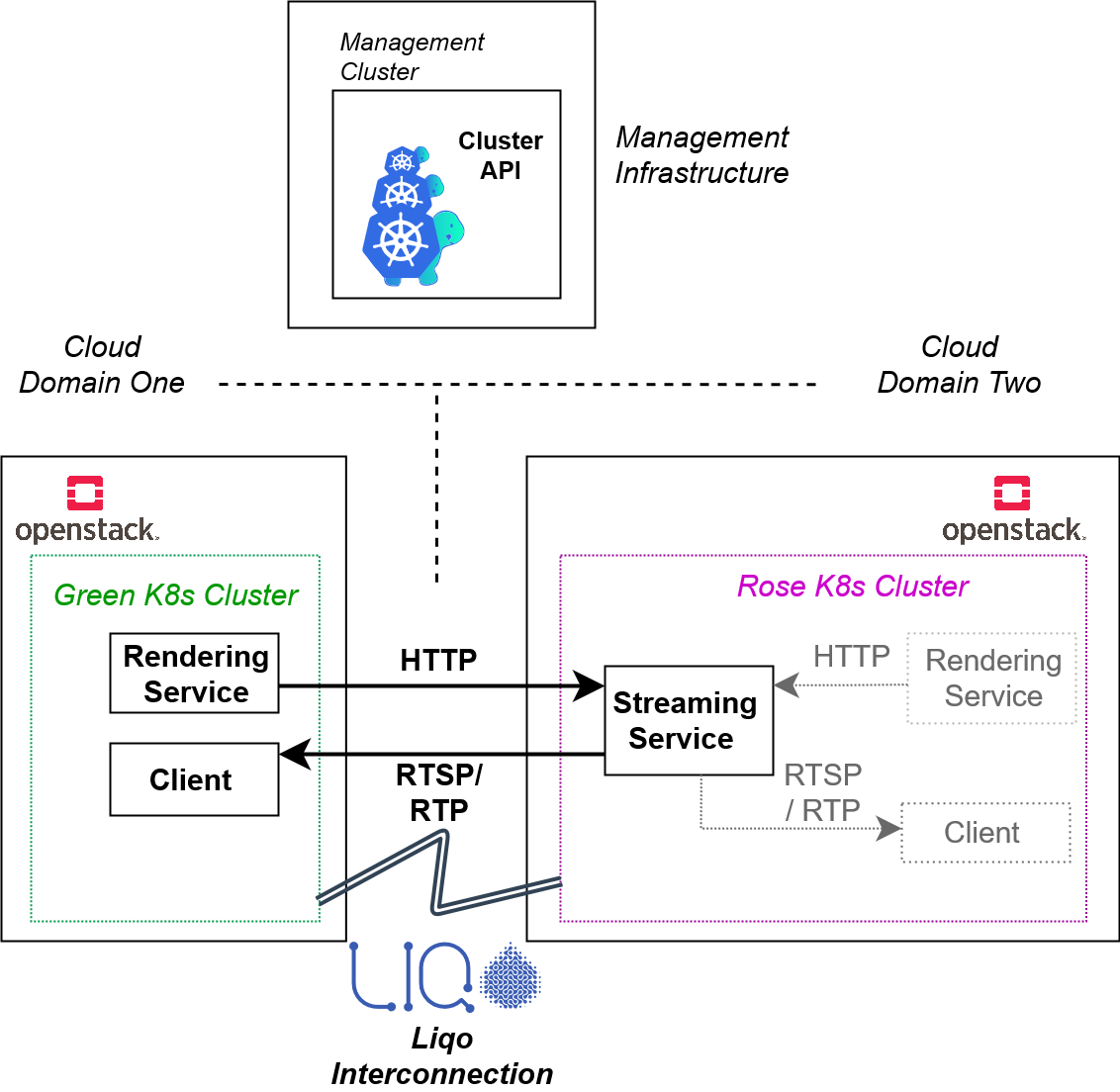} \\
    \small (b) Multi-Cluster Interconnection Scenario with Liqo. \\ \\
  \end{tabular}
  \caption{The implementation of the proposed solution in the frame of a cross-cluster video streaming use-case that includes two clusters.}
  \label{fig:experimentation_evaluation}
\end{figure}

Despite the fact that the proposed solution is constructed on the basis of a XR service agnostic manner, in the sense that it can facilitate any type of XR service, it is of paramount importance to showcase the applicability and efficiency of the proposed solutions within the frame of a specified use-case. Since Video Streaming is a quite important aspect of XR services, the authors of this work have chosen to examine a cross-cluster video streaming use-case. Fig.~\ref{fig:experimentation_evaluation} represents (a) an environment provisioned via Cluster API and (b) a Multi-Domain environment hosting a cross-cluster video streaming use-case connected using Liqo that consists of two clusters.

In the frame of the examined cross-cluster video streaming use-case, these nodes are used to host three key components. A Rendering Service (using ffmpeg\footnote{https://ffmpeg.org/}) sends a source feed using Hypertext Transfer Protocol~(HTTP) to a Streaming Service. The Streaming Service (using ffserver\footnote{https://trac.ffmpeg.org/wiki/ffserver} as a media server) and a Client~(another ffmpeg instance using Real Time Streaming Protocol~(RTSP)) to consume the video. 

\section{Experimental Evaluation \& Discussion}
\label{exp}

To showcase the validity of the proposed solution in terms of facilitating multi-cluster management \& networking, two experiments were conducted. The first study focused on assessing the speed at which various cluster sizes and distributions could be automated and provisioned using the Cluster API. Thus, it examines the effect that the use of Cluster API has on the overall latency from the perspective of provisioning times. The second study investigated the efficiency of Liqo in the context of video streaming, an important functionality in XR services, in terms of end-to-end latency and resource consumption. 

\subsection{Automated and Declarative Cluster Orchestration}

Automated provisioning through Cluster API is crucial for minimizing infrastructure and tooling bootstrapping time, especially in impractical scenarios for larger Kubernetes clusters. This evaluation centers on automating and assessing the provisioning times for various Kubernetes cluster sizes and distributions, including lightweight options like K3s, prevalent in resource-constrained environments. The objective is to scrutinize Cluster API's support for diverse Kubernetes distributions and analyze the provisioning times for each.

As depicted in Fig.~\ref{fig:experimentation_evaluation}.(a), infrastructure is conceptually divided into management and cloud, thus reflecting the two essential components of the proposed solution. The first consists of a Kubernetes cluster hosting the \textit{Cluster API}~(v1.3.1) components and a set of \textit{Cluster API providers}, namely two Control Plane providers, \textit{kubeadm}~(v1.3.1) and \textit{k3s}~(v0.1.5) and one infrastructure provider, \textit{OpenStack}~(v0.7.0). Such management represents the key elements for orchestrating the remaining edge cloud. \textit{Kubeadm} and \textit{k3s} portray two widely used control plane installation options. On the other hand, the second part of the scenario consists of the cloud infrastructure using an OpenStack~(Microstack Ussuri version) default installation to host the various Kubernetes clusters and their nodes where XR services will run (e.g., the clusters coloured in green and rose illustrated in Fig.~\ref{fig:experimentation_evaluation}.(a)).

The evaluation involved measuring the time required for generating cluster resource definitions (\textit{using Cluster API syntax}), applying these resources to the management cluster, creating corresponding Cloud resources (i.e., Virtual Machines in OpenStack), and configuring the Kubernetes cluster. \textit{Clusterctl}, part of \textit{Cluster API}, was used to generate cluster definitions, leveraging default templates from each provider. For K3s on OpenStack, where no template existed, we created one\footnote{\url{https://github.com/cluster-api-provider-k3s/cluster-api-k3s/pull/24}} specifically for Kubernetes clusters with \textit{k3s} on OpenStack. The evaluation considered one, three, and five nodes for each Kubernetes distribution, representing one control plane, one control plane and two workers, and one control plane and four workers, respectively. Each node and Virtual Machine utilized \textit{Ubuntu} images, with a \textit{m1.medium} flavor featuring 2 vCPUs, 4GB RAM, and 20GB disk for both control plane and worker nodes.

\begin{figure}[htb]
    \centering
    \includegraphics[width=\linewidth]{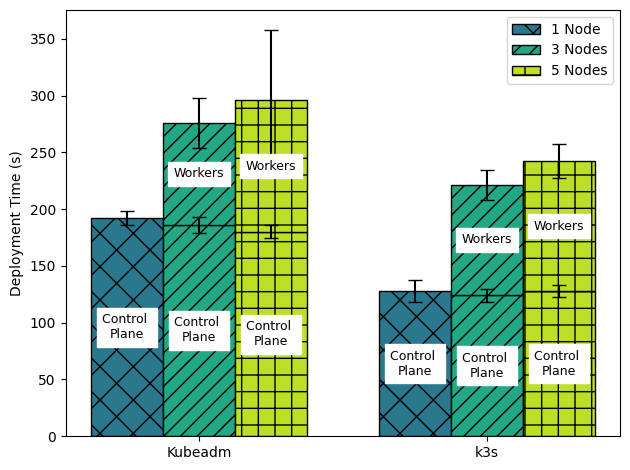}
    \caption{Cluster API deployment time by Kubernetes distribution and cluster size.}
    \label{fig:capi-times}
\end{figure}

Fig.~\ref{fig:capi-times} displays the experiment results, depicting deployment times for both the control plane and total cluster nodes. Deployment time is defined as the period until the control plane indicates \textit{Ready} for single-node clusters or until all nodes indicate \textit{Ready} for multi-node clusters. The tests were conducted 30 times, with error bars indicating the standard deviation. The results indicate no significant difference in deployment time between the two Kubernetes distributions. Total deployment times ranged from $150$ to $270$ seconds, encompassing the creation of corresponding Virtual Machines on the target infrastructure. Notably, \textit{K3s} clusters demonstrated faster provisioning than Kubernetes clusters with \textit{kubeadm} across all sizes. It is crucial to consider various factors influencing these times, including internal Cluster API reconciliation logic, cluster provisioning steps, target infrastructure, additional software installation (e.g., CNI), and checks required before marking a cluster as \textit{Ready}. Additionally, scaling the number of nodes did not proportionally increase total deployment time, suggesting simultaneous bootstrapping of the control plane and remaining nodes by Cluster API. 

\subsection{Dynamic Cross-cluster Networking}

Since video Streaming is a relevant aspect of XR services, we chose to evaluate the integration of Liqo to support cross-cluster video streaming. This evaluation aims to shed light on the effect that the use of Liqo has in terms of latency (due to overhead) and resource consumption. Towards achieving this goal, Liqo was compared against Kubernetes' NodePort. Although NodePort's approach of publicly exposing all services that need to communicate with each other prevents it from being a viable option for cross-cluster networking due to security and scalability concerns, it is capable of serving as a benchmark against Liqo in terms of the end-to-end latency and resource consumption. 

The video streaming experiment is composed of the three aforementioned key components, as depicted in Fig.~\ref{fig:experimentation_evaluation}.(b). Based on this use-case, five scenarios were devised: 1) all components deployed in the same Kubernetes cluster; 2) two local clusters on top of a singular OpenStack infrastructure connected via Liqo. In this, one cluster hosted the rendering and client components, while the other accommodates the streaming service; 3) the same two local clusters of \textbf{sce.2} but exposing services through NodePort; 4) two remote clusters, each one on top of a different OpenStack Infrastructure. One OpenStack deployment in Frankfurt and a second one in Geneva, connected through Liqo; 5) the same remote clusters of \textbf{sce.4} but exposing services through NodePort;

\begin{figure}[!tb]
    \centering
    \includegraphics[width=\linewidth]{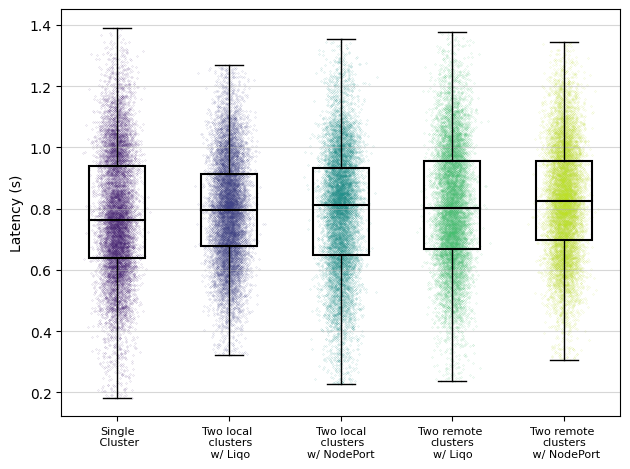}
    \caption{Latency between streaming and view times for each scenario.}
    \label{fig:latency}
\end{figure}

\begin{figure}
  \centering
  \begin{tabular}{ c @{\hspace{30pt}} c }
    \includegraphics[width=\columnwidth]{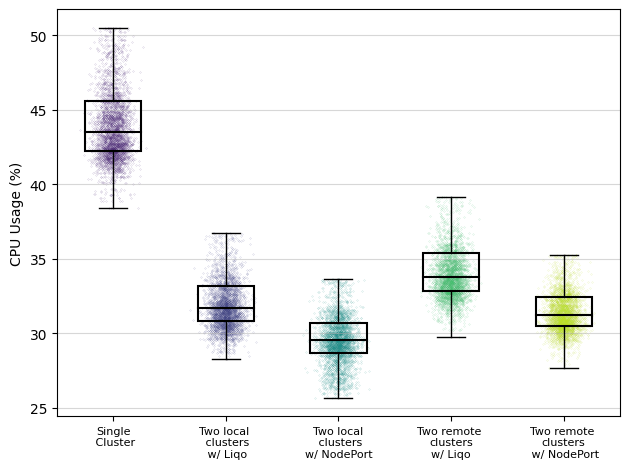} \\
    \small (a) CPU consumption. \\ \\
    \includegraphics[width=\columnwidth]{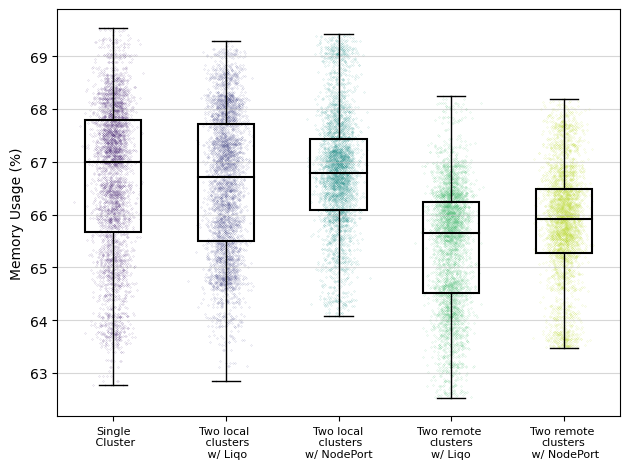} \\
    \small (b) Memory consumption. \\ \\
  \end{tabular}
  \caption{Resource Consumption for each scenario.}
  \label{fig:resource_consumption}
\end{figure}

In all scenarios, each Kubernetes cluster was configured with 2 vCPUs, 4GB of RAM, and a 20GB disk. The setup utilized Ubuntu images for deployment. To assess Liqo overhead, we measure the overall end-to-end system latency (from the moment the frame is generated to the moment it is consumed). For each scenario, we conducted five independent runs of approximately $30$ minutes of video streaming for a total of $150$ min. For measuring latency, timestamps were embedded into the frames themselves using \textit{ffmpeg} filters (generation and consumption timestamps, respectively), and later the difference was computed via Optical Character Recognition using Tesseract engine\footnote{https://github.com/tesseract-ocr/tesseract}. For each scenario over $10000$ frames were analysed. For each scenario, we also recorded the CPU and RAM consumption using Prometheus\footnote{https://prometheus.io/}. Fig.~\ref{fig:latency} shows the aggregated latency values of all runs per scenario. Whereas, Fig.~\ref{fig:resource_consumption} investigates resource consumption in the form of the CPU and Memory values. For scenarios with two clusters (\textbf{sce.2} to \textbf{sce.5}), the resource consumption values represent the average of the two. 

Although \textbf{sce.1} is not aligned with the underlying premise of the next generation of XR services will be distributed across domains (and clusters). Instead, its results can provide us with hints about the overall latency regardless of the multi-domain aspect. According to the results plotted in Fig.~\ref{fig:latency}, the $50th$ percentile for \textbf{sce.1} was equal to $762$ms, thus surpassing the other scenarios. However, in all scenarios the $90th$ percentile consistently stayed around the 1-second mark, meaning that the significant majority of frames had an end-to-end latency equal to or below this value. Furthermore, \textbf{sce.1}, despite providing lower average values, presented slightly more data dispersion (with a standard deviation of $202$ms) characterized by the occurrence of clustering around the $1$s value. These results can be explained by the fact that all the processing occurred in the same cluster. Even so, it's important to note that no bottleneck took place according to the resource consumption results depicted in Fig.~\ref{fig:resource_consumption}. Liqo performed slightly better compared to NodePort in the context of both local and remote clusters. More specifically, in \textbf{sce.2} and \textbf{sce.3}, the $50th$ percentiles were equal to $794$ms and $811$ms, while in \textbf{sce.4} and \textbf{sce.5} the $50th$ percentiles were equal to $801$ms and $824$ms. The low difference between local and remote clusters can be attributed to the relatively small physical distance among cluster locations (Geneva to Frankfurt). Although these values represent the overall end-to-end latency as perceived by a client of this particular scenario, hence, they are also dependent on the application itself, they demonstrate how Liqo's overlay network is capable of supporting distributed XR architectures in terms of end-to-end latency. 

Furthermore, the experimental results showcased in Fig.~\ref{fig:resource_consumption} indicate that the use of Liqo did not result in an prohibitively increased resource consumption in terms of CPU and memory. As expected, in \textbf{sce.1} the CPU usage was higher with a median of $43.2\%$. For the remaining scenarios, Liqo revealed relatively higher CPU consumption when compared to the NodePort. More specifically, in \textbf{sce.2} and \textbf{sce.3}, the $50th$ percentiles were equal to $31.5\%$ and $29.4\%$, while in \textbf{sce.4} and \textbf{sce.5} the $50th$ percentiles were equal to $33.5\%$ and $31.0\%$. These values can be attributed to the overlay strategy of having dedicated network tunnels between the clusters. Furthermore, the memory usage results do not indicate that the use of Liqo was accompanied by an increase in memory consumption. Instead, the examined scenarios exhibited an overall relatively similar behaviour varying approximately between $62\%$ and $70\%$, regardless of whether or not they incorporated Liqo. 

The experimental results that were explored in the frame of this section demonstrated that the proposed solution is capable of supporting not only the functional requirements of XR services that were explored in section \ref{con}, but also the aforementioned requirements of XR services in terms of latency and resource consumption.

\section{Conclusion}
\label{conclusion}
This work identified the QoS and functional requirements of XR services in multi-cluster deployments. After carefully researching and comparing the available technologies that constitute the state-of-the-art, the authors devised a solution consisting of two contemporary frameworks capable of facilitating the aforementioned requirements. These frameworks are Cluster API for multi-cluster orchestration and Liqo for multi-cluster networking. We evaluated the proposed solution through two experimental processes. The first one focused on the evaluation of the automation and provisioning times of different cluster sizes and distributions using Cluster API. The second one examined the efficiency of Liqo in the context of cross-cluster video streaming, which is a prominent use-case in the frame of XR services. The results of the experiments indicate that the proposed solution is fully capable of implementing multi-cluster deployments in a manner that facilitates the QoS and functional requirements of XR applications.

\section*{Acknowledgment}
The research leading to these results received funding from the European Union's Horizon 2020 research and innovation programme under grant agreement No 101016509 (project CHARITY). The paper reflects only the authors' views. The Commission is not responsible for any use that may be made of the information it contains.

\bibliographystyle{IEEEtran}
\bibliography{references}

\end{document}